\documentclass[prl,twocolumn,showpacs]{revtex4}
\usepackage{graphicx}
\usepackage{amsfonts}

\usepackage{amsmath, amsthm, amssymb}
\usepackage{times}
\newcommand{\ve}[1]{\mathbf{#1}}
\newcommand{\kk}{\ve{k}}

\newcommand{\vS}{\mathbf{S}}
\newcommand{\vs}{\mathbf{s}}

\newcommand{\vsig}{\boldsymbol{\sigma}}
\newcommand{\ud}{\mathrm{d}}

\newcommand{\beq}{\begin{equation}}
\newcommand{\eeq}{\end{equation}}

\begin{document}
\title{Kondo resonance narrowing in $d$- and $f$-electron systems}
\author{Andriy H. Nevidomskyy}
\email{nevidomskyy@cantab.net}
\author{P. Coleman}
\affiliation{Department of Physics and Astronomy,
Rutgers University, Piscataway, N.J. 08854, USA}

\date{\today}
\begin{abstract}
By developing a simple scaling theory for the effect of Hund's
interactions on the Kondo effect, we show how an exponential narrowing
of the Kondo resonance develops in magnetic ions with large Hund's
interaction.  Our theory predicts
an exponential reduction of the Kondo temperature with spin S of the
Hund's coupled moment, a little-known effect first observed in
d-electron alloys in the 1960's, and more recently encountered in 
numerical calculations on multi-band Hubbard models with Hund's
interactions.  We discuss the consequences of Kondo resonance
narrowing for the Mott transition in d-band materials, particularly
iron pnictides, and the
narrow ESR linewidth recently observed in ferromagnetically
correlated f-electron materials.
\end{abstract}


\maketitle



\begin{figure}[!tb]
\begin{center}
{\includegraphics[width=0.48\textwidth]{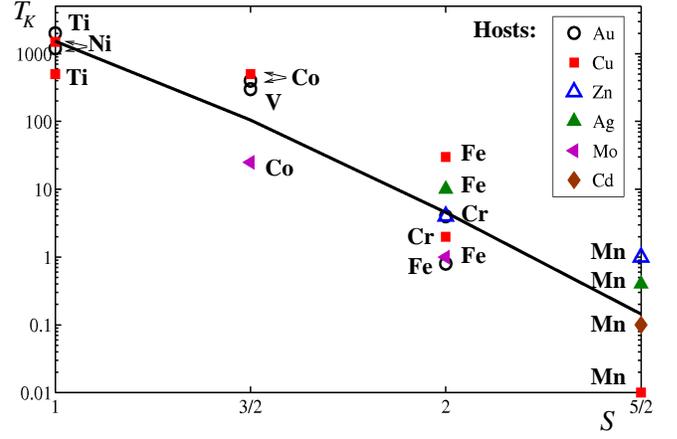}}
\end{center}
\caption{Measured  values \cite{Daybell-73} 
of the Kondo temperature $T_K^*$ in host alloys Au, Cu, Zn, Ag, Mo, and Cd
containing transition metal impurities, plotted
  vs. the nominal size $S$ of the spin. Solid line is the fit
  to Eq.~(\ref{theformula}) with $\Lambda_0\equiv J_HS$.
 } 
\label{Fig.Tk-fit}
\end{figure}

The theory of the Kondo effect forms a cornerstone in current
understanding of correlated electron systems~\cite{Hewson-book}.
More than forty
years ago, experiments  on $d$-electron materials found that 
the characteristic  scale of spin fluctuations of magnetic impurities,
known as the Kondo temperature, narrows exponentially with the size $S$ of the 
impurity spin~\cite{Daybell-RMP68} 
(Fig \ref{Fig.Tk-fit}). 
An explanation of this effect was proposed~\cite{Daybell-RMP68} 
based  on an early theory of the 
Kondo effect by Schrieffer \cite{Schrieffer67}, who 
found that strong Hund's coupling 
leads to a $2S$-fold reduction of the Kondo coupling constant. Surprisingly, 
interest in this phenomenon waned 
after the 1960's.
Motivated by a recent resurgence of interest 
in $f$- and $d$-electron systems, especially 
quantum critical heavy electron systems~\cite{qcriticalrev}, and
pnictide superconductors~\cite{pnictiderev},  
this paper revisits this little-known phenomenon,  which we refer to 
as  ``Kondo resonance narrowing'', placing it in a modern context. 

The consequences of Kondo resonance narrowing have recently 
been re-discovered in calculations on multi-orbital
Hubbard and Anderson models~\cite{bulla-Hunds, millis-3orbital}.
Numerical renormalization group studies found that the introduction of 
Hund's coupling into the Anderson model causes an 
exponential reduction in the Kondo temperature~\cite{bulla-Hunds}. 
The importance of Hund's effect
has also arisen in the context of iron pnictide superconductors
\cite{haule-La1111, haule-Hunds}, where it appears to play a key role
in the development of ``bad metal'' state in which the $d$-moments
remain unquenched down to low temperatures.

In this paper, we show that Kondo resonance narrowing 
can be simply understood within a scaling theory 
description of the 
multi-channel Kondo model with Hund's interaction. 
The main result is an 
exponential decrease of the Kondo temperature that develops when 
localized electrons lock together to form a large spin $S$,
given by the formula
\begin{equation}\label{theformula}
\ln  T_{K}^{*} (S)=\ln  \Lambda_0 - (2S)\ln \left(\frac{\Lambda_{0}}{T_{K}} \right).
\end{equation}
Here, $T_{K}$ is the ``bare'' spin $1/2$ 
Kondo temperature and $\Lambda_{0} = \hbox{min} ( J_{H}S, U+E_{d},
|E_{d}|)$ is the scale at which the locked spin $S$ develops under the
influence of a Hund's coupling, while $U$ and $E_{d}$ are the
interaction strength and position of the bare $d$-level.
Although this result is implicitly contained in the early works of 
Schrieffer~\cite{Schrieffer67} and Hirst~\cite{Hirst71}, a detailed
treatment has to our knowledge, not previously been given. 



To develop our theory, we consider $K$ spin $1/2$ impurity spins at a
single site, ferromagnetically interacting 
via Hund's coupling $J_H$, each coupled 
to a conduction electron channel of band-width $D$ 
via an antiferromagnetic
interaction $J$:
\beq
H = \sum\limits_{\kk,\sigma,\mu} \varepsilon_k
c_{\kk\sigma \mu}^\dag c_{\kk\sigma \mu}   -J_H\left(\sum\limits_{\mu=1}^K
\vs_\mu\right)^2 + J\! \sum\limits_{\mu=1}^K \vs_\mu\cdot\vsig_\mu, \label{H0}
\eeq
where $\varepsilon_k$ is the conduction electron energy, $\mu= 1,K$
is the channel index and 
$\vsig_\mu=\sum_\kk c_{\kk\alpha \mu}^\dag \vsig_{\alpha\beta}\, c_{\kk\beta \mu}$ is
the conduction electron spin density in channel $\mu$ at the origin.
We implicitly assume
that Hund's scale $KJ_{H}$ is smaller than $D$. 
When derived from an Anderson
model of $K$ spin-$1/2$ impurities, then $D = \hbox{min}
(E_{d}+U,\vert E_{d}\vert )$ is the cross-over scale at which local
moments form while 
$J = |V_{k_F}|^2 (1/(E_d+U) + 1/|E_d |)$is the Schrieffer--Wolff form
for the Kondo coupling constant~\cite{Hewson-book},
where $V_{k_F}$ is the Anderson hybridization averaged over the Fermi
surface and $E_{d}<0$ is negative. 

The behavior of this model 
is well understood in the two extreme
limits~\cite{nozieres-blandin}: for $J_H\!=\!\infty $,  the $K$
spins lock together,  forming a $K$-channel
spin $S\!=\!K/2$ Kondo model. The opposite limit $J_H=0$
describes $K$ replicas of the 
spin-$1/2$ Kondo model. 
Paradoxically,  the leading exponential dependence  of the  
Kondo temperature on the coupling constant 
$T_K\sim D e^{-1/2J\rho}$ 
in these two limits is independent of the size of the spin.
However, as we shall see in the cross-over between the two limits, the
projection of the Hamiltonian into the space of
maximum spin leads to a $(2S)$-fold reduction in the
Kondo coupling constant.

\begin{figure}[!tb]
\begin{center}
{\includegraphics[width=0.46\textwidth]{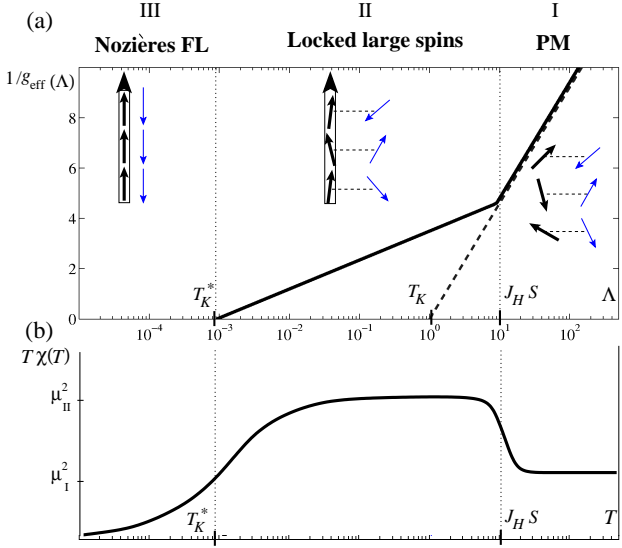}}
\end{center}
\caption{(a) Schematic showing the behavior of
the running coupling constant $g_\text{eff}(\Lambda)
=J(\Lambda)\rho K_\text{eff}$ on a logarithmic 
scale, with $K_\text{eff}$ the effective number of conduction electron
channels per impurity spin ($K_\text{eff}=1$ in region I and $K$ in
regions II and III). (b) Schematic showing 
effective moment  $\mu_\text{eff}^2(T) = T \chi (T)$ in terms of the 
susceptibility $\chi(T)$, showing the enhancement (\ref{ratio-Meff})
in region II and the loss of localized moments due to Kondo screening in region III.}
\label{Fig.scaling}
\end{figure}

We now study the properties of this model as a
function of energy scale or cut-off $\Lambda$. 
Qualitatively, we expect 
three distinct regions  depicted in Fig.~\ref{Fig.scaling}(a):\\
(I) $\Lambda\gg J_H S$: a spin-$1/2$ disordered paramagnet
characterized by a high temperature Curie 
magnetic susceptibility 
\beq
\chi_\text{I}(T) = K\, \frac{(3/4)(g\mu_B)^2}{3k_BT} \label{chi-I}
\eeq
(where $g$ is the g-factor of the electron), 
with effective moment $(\mu_{\text{eff}}^\text{I})^2= 3 K/4$;\\
(II) $T_K^*\ll\Lambda\ll J_H S$: an unscreened big spin $S=K/2$ is formed
above an emergent Kondo energy scale $T_K^*$; \\
(III) $\Lambda\ll T_K^*$: the Nozi\`eres Fermi liquid ground state of
the $K$-channel $S=K/2$ Kondo problem.

\begin{figure}[!tb]
\begin{center}
{\includegraphics[width=0.42\textwidth]{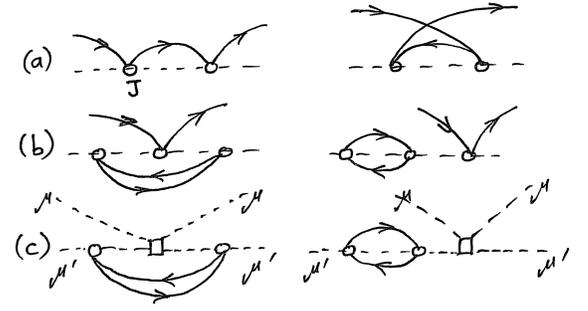}}
\end{center}
\caption{The diagrams appearing in (a) one-loop and (b) two-loop
  RG equations for Kondo coupling J (open circles), with solid lines
  denoting the conduction electron propagators and dashed line -- the
  impurity spin. (c) The lowest order diagrams in the RG flow of
  Hund's coupling (a square vertex denotes bare $J_H$).}
\label{Fig.diagram}
\end{figure}

We employ the ``Poor Man's scaling'' approach~
\cite{Anderson70,
Hewson-book}, in which the leading renormalization flows are followed
as the electrons in the conduction band are systematically decimated
from the Hilbert space. 
By computing the diagrams depicted in Fig.~\ref{Fig.diagram}, we
obtain the following renormalization group (RG) equations in region I:
\vspace{-1mm}
\begin{eqnarray}
\text{(I):}\qquad \frac{\ud\, (J\!\rho)}{\ud \ln \Lambda} & = & -2\, (J\rho)^2 +
2 \,(J\rho)^3 \label{Jscaling} \\
\frac{\ud\, (J_H\rho)}{\ud \ln \Lambda} & = & 4 (J\rho)^2 J_H\rho 
\label{JHscaling}
\end{eqnarray}
where $\rho$ is the density of states of the conduction electrons at the
Fermi level. The first equation is the well-known beta function for
the Kondo model, which to this order is independent of Hund's 
coupling. As we decimate the states of the conduction sea, 
reducing the band-width $\Lambda$  
down to the Hund's scale $J_HS$, to leading logarithmic order we obtain
\vspace{-1mm}
\beq
\rho
J(\Lambda)=\left. \frac{1}{\ln\left(\frac{\Lambda}{T_K}\right)}\right|_{\Lambda=J_H S}
\equiv\rho J_\text{I}.  \label{J1}
\eeq
There is a weak 
downward-renormalization of the
Hund's coupling $J_H$ described by Eq.~(\ref{JHscaling}), which
originates in the two-loop diagrams (Fig.~\ref{Fig.diagram}c)). 
In the leading logarithmic approximation, we may approximate 
$J_H$ by a constant.

Once $\Lambda$ is reduced below $J_{H}S$, the individual local moments become 
locked into a spin $S=K/2$, similar to the effect discussed in
Ref.~\onlinecite{2impurity} for the case of two impurities coupled by
ferromagnetic RKKY interaction.  
The low-energy properties of the system in region II are described by
a Kondo model of spin $K/2$ with $K$ conduction
electron channels: 
\vspace{-2mm}
\beq
H_{\text{eff}}^\text{II} = J^*(\Lambda) \sum_{\mu=1}^K \vS \cdot \vsig_\mu, \label{Heff}
\eeq
To obtain the value of $J^{*}$, we must project the original model
into the subspace of maximum spin $S$.
By the Wigner-Eckart theorem, the matrix element of any vector
operator acting in the basis of states $|S_z\rangle$ of big spin
$S=K/2$ is related by a constant prefactor to 
$\vS$ itself, i.e.
\vspace{-1mm}
\beq
\langle S S_z|\vs_\mu|S S_z \rangle = g_S \langle S_z|\vS|S_z \rangle \label{Wigner}
\eeq
Summing both sides of the equation over impurity index $\mu=1,\dots K$,
one obtains $\langle S S_z|\sum_\mu\vs_\mu|S S_z \rangle = g_S\,K\,
S_z$. However since $\sum_\mu \vs_\mu = K\vs \equiv \vS$, one arrives at the
conclusion that $g_S K=1$, therefore determining the value of the constant coefficient $g_S=1/K$ in Eq.~(\ref{Wigner}). 
Comparing Eqs.~(\ref{H0}) and (\ref{Heff}), we arrive at the effective
Kondo coupling: 
\beq
J^* = J/K. \label{J_K*}
\eeq
This equation captures the key effect of crossover from region I to
region II in Fig.~\ref{Fig.scaling}. This result was first 
derived in the early work 
on the multi-channel Kondo problem by Schrieffer~\cite{Schrieffer67},
where the limit of $J_H\to \infty$ was  implicitly assumed, and also
appears for the particular case of $K=2$ in the study of the two-impurity Kondo
problem by Jayaprakash \emph{et al.}~\cite{2impurity}.

To one loop order, the scaling 
equation for $J^*(\Lambda)$ in region II 
is identical   to that of  region I
(\ref{Jscaling}), namely
$\ud\,(J^*\rho)/\ud\ln\Lambda\approx -2(J^*\rho)^2$, though its size
is $K$ times smaller. 
To avoid the
discontinuous jump in coupling constant at the crossover, 
it is more convenient to consider
$g_\text{eff}\equiv J(\Lambda)\rho K_\text{eff}$,  where the effective
number of channels $K_{eff}=1$
and $K$ in regions I and II respectively.
This continuous variable  satisfies
\beq
\text{(II):}\qquad \frac{\ud\, g_\text{eff}}{\ud \ln \Lambda}  =  -\frac{2}{K}\, g_\text{eff}^2 +
\frac{2}{K}\,g_\text{eff}^3. \label{J*scaling}
\eeq
so the speed at which it scales to strong coupling becomes
$K$ times smaller in region II (see Fig.~\ref{Fig.scaling}a).
Solving this RG equation to leading order, and setting 
$g_\text{eff} (\Lambda=T_{K}^{*})\sim 1$, we obtain 
$T_K^*\sim (J_H S)\left(D/J_HS\right)^K e^{-\frac{K}{2J\rho}}$ for the
renormalized Kondo scale.
Comparing this with the bare Kondo scale $T_K\sim
D e^{-1/2J\rho}$, we deduce
\beq
T_K^*\sim J_H S \left(\frac{T_K}{J_HS}\right)^K\equiv T_K \left(\frac{T_K}{J_HS}\right)^{K-1}\label{Tk*},
\eeq
from which formula (\ref{theformula}) follows.
This exponential suppression of the
spin tunneling rate can be understood as a result of a $2S$-fold
increase in the classical action associated with a spin-flip. 


These results are slightly modified when 
the two-loop terms in the scaling 
are taken into account. 
The expression for $T_K$ now acquires a pre-factor, 
$T_K=D\sqrt{J\rho}\,e^{-1/2J\rho}$ and $J_{H}$ is weakly
renormalized so that 
\beq
T_K^* = (\tilde{J}_HS)  \left(\frac{T_K}{\sqrt{K} \tilde{J}_HS} \right)^K, \label{Tk*2}
\eeq
where $\tilde{J}_H$ is determined from the 
quadratic equation
\beq
x^2-x\left(x_0+\frac{4}{\ln(D/T_K)}  \right)+4=0.
\eeq
where $x=\ln(\tilde{J}_HS/T_K)$ and
$x_0\equiv \ln(J_HS/T_K)$. 

The magnetic impurity susceptibility 
in region II becomes
\beq
\chi_{\text{imp}}^* \!=\! \frac{(g\mu_B)^2}{3
  k_BT} S(S+1)\left( 1\! - \!\frac{1}{\ln\left(\frac{T}{T_K^*}\right)} 
\!+\!\mathcal{O}\!\left( \frac{1}{\ln^2\!\left(\frac{T}{T_K^*}\right)}
  \right)\! \right) \label{chi-II},
\eeq
from which we see that the enhancement of the magnetic moment at the
crossover is given by  (see  Fig.~\ref{Fig.scaling}b))
\beq
\left(\frac{\mu_\text{eff}^\text{II}}{\mu_\text{eff}^\text{I}}\right)^2
= \frac{K+2}{3}. \label{ratio-Meff}
\eeq



When the temperature is ultimately reduced below the exponentially
suppressed  Kondo scale
$T_K^*$,  
the big spins $S$  become screened to form a
Nozi\`eres Fermi liquid
~\cite{nozieres1}.  A phase-shift description of the
Fermi liquid predicts that~\cite{yoshimori,nozieres-blandin} 
the Wilson ratio $W \equiv
 \frac{\chi_{\text{imp}}}
{\chi_0}/\frac{\gamma_{\text{imp}}}{\gamma_{0}}$ is given
by
\beq
W_K= \frac{2(K+2)}{3} \equiv 2 \left(\frac{\mu_\text{eff}^\text{II}}{\mu_\text{eff}^\text{I}}\right)^2, \label{W_K}
\eeq
which, compared with the classic result $W_{1}=\!2$ for the one-channel
model~\cite{nozieres1}, contains a factor of the moment
enhancement. This result holds in the extreme limit $J_{H}\!\gg\!  T_{K}$. 
More generally, $W$
depends on the ratio $\eta=U^*/J_H^*$
of a channel-conserving interaction $U^{*}$
to an  inter-channel Hund's coupl\-ing $J_{H}^{*}$ in the Fermi liquid
phase-shift analysis, giving rise to 
\beq
W_K(\eta)=2\left(1+\frac{K-1}{2(1+\eta)+1}\right). \label{Wilson}
\eeq 
On general grounds we expect $\eta\sim T_K/J_H$.





We end with a discussion of the broader implications 
of Kondo resonance narrowing for $d$- and $f$-electron materials. 
This phenomenon provides a simple explanation of the drastic
reductions in spin fluctuation scale observed in the classic
experiments of the sixties \cite{Daybell-RMP68}. Our treatment brings
out the important role of Hund's coupling in this process. 
One of the untested predictions of this theory is a
linear rise of the Wilson ratio $W$ with spin $S\!=\!K/2$
(\ref{W_K}), from a value $W[1]=$2.7 in Ti and Ni, to $W[5/2]=$4.7
in Mn impurities.
Taking together with
the early data, Fig.~\ref{Fig.Tk-fit}, we are able to essentially
confirm the early speculation \cite{Schrieffer67} 
that were Hund's 
coupling absent, the Kondo effect would take place at such high
temperatures that dilute $d$-electron 
magnetic moments would be unobservable.
This is, in essence, the situation 
for Ti impurities in gold, where the Kondo temperature of $S=1$ 
moments becomes so high that magnetic behavior is absent 
below the melting temperature of gold.
On the other hand, the Kondo resonance narrowing effects of Hund's interaction 
can become so severe, that the re-entry
from region II into the quenched Fermi liquid becomes too low to observe.
This is the case for $S=5/2$ Mn in gold, where $T_{K}^{*}$
is so low that it has never been observed; the
recent observation of a ``spin
frozen phase'' in  DMFT studies~\cite{millis-3orbital} may be a 
numerical counterpart.

What then,  are the possible implications  for
dense $d$-electron systems? 
In those materials,
the ratio of Kondo temperature to the Hund's coupling 
will be strongly dependent on structure, screening and chemistry. 
In cases where  $J_{H}\!\ll\! T_{K}$, the physics of
localized magnetic moments will be lost and the $d$-electrons will 
be intinerant. 
On the other hand, the situation where
$J_{H}\!\gg\! T_{K}$ will almost cer\-tain\-ly lead to long range magnetic
order with localized $d$-elec\-t\-rons.
Thus in multi-band systems, the cri\-te\-rion $J_{H}\sim T_{K}$
determines the boundary between localized and itinerant behavior,
playing the
same role as the condition $U/D\sim 1$ in one-band Mott insulators. 

These issues may be of particular importance 
to the ongoing debate about the strength of electron correlations
in the FeAs family of high-temperature
superconductors~\cite{pnictiderev, pnictides-JACS,
  pnictides-chen, pnictides-delaCruz}. 
Current wisdom argues that in a multi-band system, the
critical interaction $U_c$ necessary for the Mott metal-insulator
transition grows linearly with the number of bands
$N$~\cite{florens02, ono03}, favoring
a viewpoint that iron pnictide materials are itinerant 
metals lying far from the Mott regime. 

In essence, Hund's coupling converts a one channel Kondo model to a
$K$-channel model (\ref{Heff}). 
Large-$N$ treatments of these models  
show that the relevant control parameter is 
the ratio $K/N$~\cite{lebanon-coleman}, rather than $1/N$.
By repeating the large-$N$
argument of Florens \emph{et al.}~\cite{florens02}, we conclude
that
the critical value of the on-site interaction $U$ for the Mott
transition is 
\beq
U_c \propto \left(N/K\right) V_{k_F}^2\rho.
\eeq 
Thus Hund's coupling compensates for multi-band behavior, restoring 
$U_c$ to a value comparable with one-band models.
Recent DMFT calculations on the two-orbital Hubbard 
model~\cite{bulla-Hunds} support this view, finding that $U_c$
is reduced from $U_c\approx 3D$ to
$U_c^*\approx 1.1D$ when $J_H/U=1/4$. 

LDA+DMFT studies of iron pnictide materials
\cite{haule-Hunds} 
conclude that in order to reproduce the incoherent bad metal features
of the normal state, a value
of $J_H\!\sim\!0.4$~eV is required, resulting in $T_K^*$ of
the order of 200~K. Fitting this with Eq.~(\ref{Tk*}) results in a nominal
$T_K\sim3000$~K and a ratio $T_K/J_HS\sim0.4$. By contrast, for 
dilute Fe impurities in
Cu~\cite{Daybell-73} $T_K^*\approx\!20$~K, from
which we extract a bare ratio $T_K/J_HS\sim0.2$ and
$T_K\!\sim\!3500$~K.
The bare Kondo temperature is essentially the same in
both cases, but $T_K/J_HS$ is significantly 
increased due to screening of $J_H$ in the iron pnictides, placing them
more or less at the crossover $J_H\sim T_K$. 
A further sign of
strong correlations in iron pnictides derives from the Wilson ratio,
known to be  $\sim$1.8 in SmFeAsO~\cite{cimberle09} and about
4--5 in FeCrAs~\cite{wu09}, whereas Eq.~(\ref{W_K}) would predict $W\!=\!4$.


Finally, we discuss heavy $f$-electron materials, which lie 
at  the crossover between 
localized and itinerant behavior
\cite{colemanreview}. 
In these materials, spin
orbit and crystal field interactions dominate over 
Hund's interaction.  In fact, 
crystal fields are also known to 
suppress the Kondo temperature in
$f$-electron systems~\cite{CF-reducedTk}, 
but the suppression mechanism differs, involving
a reduction in the spin symmetry rather a
projective renormalization of the coupling
constant.
But the main reason that Hund's coupling is unimportant at the
single-ion level in heavy $f$-electron materials, is because 
most of them involve one $f$-electron (e.g. Ce) or one
$f$-hole in a filled $f$-shell (Yb, Pu), for which Hund's interactions are absent.

Perhaps the most interesting application of Kondo resonance narrowing to
$f$-electron systems is in the context of intersite interactions.
Indeed,  (\ref{H0})  may serve as a useful model 
for a subset of ferromagnetically correlated $f$-electron materials,
such as CeRuPO~\cite{CeRuPO}, where 
$J_H$ would characterize the scale of ferromagnetic
RKKY interactions between moments, as in Ref.~\onlinecite{2impurity}. 
In these systems, our
model predicts the formation of microscopic clusters of spins
which remain unscreened in region II down to an exponentially
small scale $\sim T_K^*$.
This exponential narrowing of the Kondo scale may provide a clue to 
the observation~\cite{fmESR08,YRS-ESR} of very 
narrow ESR absorption lines in a number of Yb and Ce heavy
fermion compounds with enhanced Wilson ratios. In particular, our theory would
predict that the Knight shift of the electron g-factor in region II is
proportional to the running coupling constant
$K (T) \propto g_\text{eff}(T) \sim 1/\ln(T/T_K^*)$, where $T_{K}^{*}$ is
the resonance-narrowed  Kondo temperature. 
A detailed study of the ESR lineshape in this context will be a subject of future work.



We would
like to acknowledge discussions with Elihu Abrahams, Natan Andrei,
Kristjan Haule, 
Gabriel Kotliar and Andrew Millis in connection with this work. 
This research was supported by NSF grant no. DMR~0605935. 

\vspace{-5mm} 
\bibliographystyle{apsrev}

\end{document}